# The genetic code degeneracy and the amino acids chemical composition are connected[†]

Tidjani Négadi[*]
Physics Department, Faculty of Science, Oran University, 31100, Es-Sénia, Oran, Algeria

Abstract
We show that our recently published Arithmetic Model of the genetic code based on Gödel Encoding is robust against symmetry transformations, specially Rumer's one U ↔ G, A ↔ C, and constitutes a link between the degeneracy structure and the chemical composition of the 20 canonical amino acids. As a result, several remarkable atomic patterns involving hydrogen, carbon, nucleon and atom numbers are derived. This study has no obvious practical application(s) but could, we hope, add some new knowledge concerning the physico-mathematical structrure of the genetic code.

Introduction
The genetic code is the "dictionary" of life that translates the encoded informational messages stored in DNA (the genes) into proteins, the constituents of the living matter. This translation process occurs in specialized organelles located in the cells, the ribosomes, where finite strings of mRNA-codons of variable length (copies of individual genes) are transformed into finite strings of amino acids, to be further processed for use. Historically, the genetic code was experimentally and completely deciphered by the middle sixtees of the past Century (after an intense international effort involving many people and ultimately leading to the Nobel Prize in 1968) by assigning to each one of the 64 ($4^3$) possible mRNA-codons, in the form XYZ where X, Y and Z are taken from the four nucleobases uracil (U), cytosine (C), adenine (A) and guanine (G), one of 20 canonical amino acids, listed in Table 2, and stop (or termination) codons. In the case of the standard genetic code, shared by the great majority of living creatures from bacteria to Humans, there are three stop codons and therefore 61 meaningful (or sense) codons coding for amino acids. The mapping 61→20 is not one-to-one as groups of codons could be associated to a single amino acid and one speaks about degeneracy and multiplet structure. For example the (amino acids) singlets are coded by one codon, the doublets by two, the triplets by three and so on. The multiplet structure obtained from experiment is shown in the table below (Table 1). There are 5 quartets (glycine, alanine, proline, threonine, valine), 9 doublets (cysteine, asparagine, aspartic acid, lysine, glutamine, glutamic acid, histidine, phenylalanine, tyrosine), 3 sextets (serine, leucine, arginine), 1 triplet (isoleucine), 2 singlets (methionine, tryptophane) and finally 3 stops. The degeneracy of a given multiplet (quartet, doublet, etc.) is defined as the number of codons in the multiplet minus 1. From the above numbers we conclude that the total degeneracy is 41 (61-20). Recently (Négadi 2007, 2008), we have constructed an arithmetic model of the genetic code, capable to reproduce its degeneracy, where the amino acids are "identified" to the decimal digits of the Gödel number associated to a specific sequence representing the total degeneracy. In this paper a more flexible setting is presented in connection with symmetry considerations. As two main new results in this contribution to NeuroQuantology, we present, in a first section, this last aspect and next, in a second one, we show that the degeneracy of the genetic code is deeply connected to the chemical composition of the 20 amino acids which, in turn, is also connected to our arithmetic model relying on the mathematical technique of Gödel encoding. As in our previous recent papers, we shall use again some few and simple arithmetic functions to reveal interesting genetic information. Two of these functions called $a_0(n)$ and $a_1(n)$ and defined for any integer n. The former gives the sum of the prime factors of n, from its prime factorization via the fundamental theorem of arithmetic, counting multiplicities, and the latter gives also the sum of the prime factors but ignoring multiplicities. Another one, $\Omega(n)$, called the Big Omega Function, gives the number of prime factors of n, counting multiplicities. For example, for n=12= $2^2 \times 3$ $a_0(12)=7$, $a_1(12)=5$ and $\Omega(12)=3$. Before proceeding, a word about how to "view" the amino acids is in order. Long time ago, Gavaudan (Gavaudan, 1971) suggested a shift of the "magic" number 20 to 23 by including the three stops in the scheme (20 amino acids and 3 stops). We have indicated recently (Négadi, 2008) another complementary, but different, reason also supporting such a shift. In our approach, we consider the degeneracy at the first-base position which is associated to the existence of the sextets and enlarge the set of 20 amino acids (aas) to the set of 23 amino acids signals (AASs) defined as follows: 17 amino acids other than the sextets are "singlets" at the first-base position and the 3 sextets S, L and R which are "doublets" at the first-base position, become the 6 new entities $S^{II}$, $S^{IV}$, $L^{II}$, $L^{IV}$, $R^{II}$ and $R^{IV}$. We shall return to this last aspect below.

---
[†]To the memory of Pierre Gavaudan who died in January 1985, 23 years ago.
[*] e-mail: negadi_tidjani@univ-oran.dz; tnegadi@gmail.com

Symmetry and Gödel encoding
Let us consider now the genetic code table (Table 1) and the four quadrants $U^{(1)}$, $C^{(1)}$, $A^{(1)}$ and $G^{(1)}$. In each one of them there are 16 codons all having the corresponding letter as the first base. For example in $U^{(1)}$ all codons have U as the first base. The table is in fact the representation of a 8×8 codon-matrix derived by us, at the very begining, from Quantum Mechanics and special solutions of the Schrödinger equation of the hydrogen atom, see our paper (Négadi, 2001) for the construction, and has interesting symmetries which we shall outline and use here. The total degeneracy and number of AASs in each one of the four quadrants are respectively as follows:

$U^{(1)}$: 8, 6 (F, $L^{II}$, $S^{IV}$, Y, C, W)
$C^{(1)}$: 11, 5 (P, $L^{IV}$, $R^{IV}$, H, Q)     (1)
$A^{(1)}$: 11, 7 (T, $S^{II}$, $R^{II}$, I, M, N, K)
$G^{(1)}$: 11, 5 (V, A, G, D, E)

Note that the singlets methionine M and tryptophane W have zero degeneracy. As our first main result, we assert that the following (degeneracy) sequence S

$$S=[19; 9, 4, 3, 2, 1, 1, 1, 1] \quad (2)$$

describes, at the same time, the following two partitions of the genetic code table into two moieties each (i) $U^{(1)}+C^{(1)}/A^{(1)}+G^{(1)}$, on the one hand, and (ii) $U^{(1)}+G^{(1)}/A^{(1)}+C^{(1)}$, on the other, and therefore both lead to the same Gödel encoding, also called Gödel numbering (see, for example, Nagel and Newman, 1971, for its historical use by Kurt Gödel in mathematical logic). As a practical definition, we recall that given a sequence $[x_1,x_2,x_3,…,x_n]$ of positive integers, the Gödel encoding of the sequence is the product of the first n primes raised to their corresponding values in the sequence. We have therefore from Eq.(2)

$$Enc(S)=2^{19}\times 3^9 \times 5^4 \times 7^3 \times 11^2 \times 13 \times 17 \times 19 \times 23 \quad (3)$$

This Gödel number is seen to be equal to 23! and is written below also in the decimal representation

$$Enc(S)=23!=25852016738884976640000 \quad (4)$$

| | | | | $U^{(1)}$ | | | $C^{(1)}$ |
|---|---|---|---|---|---|---|---|
| UUU F | UUC F | UCU S | UCC S | CUU L | CUC L | CCU P | CCC P |
| UUA L | UUG L | UCA S | UCG S | CUA L | CUG L | CCA P | CCG P |
| UAU Y | UAC Y | UGU C | UGC C | CAU H | CAC H | CGU R | CGC R |
| UAA stop | UAG stop | UGA stop | UGG W | CAA Q | CAG Q | CGA R | CGG R |
| AUU I | AUC I | ACU T | ACC T | GUU V | GUC V | GCU A | GCC A |
| AUA I | AUG M | ACA T | ACG T | GUA V | GUG V | GCA A | GCG A |
| AAU N | AAC N | AGU S | AGC S | GAU D | GAC D | GGU G | GGC G |

| AAA | AAG | AGA | AGG | GAA | GAG | GGA | GGG |
|-----|-----|-----|-----|-----|-----|-----|-----|
| K | K | R | R | E | E | G | G |

$A^{(1)}$                $G^{(1)}$

Table 1. The standard genetic code table

As a matter of fact consider, first, the case (i). Here we could describe $U^{(1)}+C^{(1)}$ having 19 degenerate codons, collectively, by the term $2^{19}$ in Eq.(3) and $A^{(1)}+G^{(1)}$ having 22 degenerate codons, in some detail, by the remaining terms describing the following groups V+A+G (9), $S^{II}+R^{II}$ (4), T (3), I (2) and D+E+N+K (1+1+1+1), where we indicated, in parenthesis, the degenaracies or partial degeneracies (the exponents). Conversely and considering a permutation in (3) which does not change the numerical value, we could describe $A^{(1)}+G^{(1)}$ having 22=19+3 degenerate codons, collectively, by the term $2^{19} \times 7^3$, and $U^{(1)}+C^{(1)}$ having 19 degenerate codons, in some detail, by the remaining terms describing the groups $L^{IV}+P+R^{IV}$ (9), $S^{IV}+F$ (4), $L^{II}$ (2) and H+Q+C+Y (1+1+1+1), where as above we have indicated the degeneracies. This is only what we considered in our last paper (Négadi, 2008). Consider now the case (ii). It is here where symmetry intervenes. We have established some years ago (Négadi 2004) the existence of a Dihedral group symmetry, $D_8$, with eight elements $R_i$ (1,2,...,8) defined by matrices acting on the codon-matrix mentioned above. $R_1$, $R_3$, $R_6$ and $R_8$ form a Klein sub-group of $D_8$ and conserve (globally) $U^{(1)}+C^{(1)}$ and $A^{(1)}+G^{(1)}$, suitable for case (i). $R_7$ performs the Rumer transformation U↔G, A↔C (Rumer, 1966) at all three base-positions in the codons with the result $U^{(1)}$↔$G^{(1)}$, $A^{(1)}$↔$C^{(1)}$, suitable for case (ii). Finally $R_2$, $R_4$, $R_5$ conserve (globally) the transformed quadrants: $G^{(1)}+A^{(1)}$ and $C^{(1)}+U^{(1)}$, again suitable for case (i). It is not difficult to see that, in this transformation process, the sequence (2) is invariant and describes both cases (i) and (ii). Let us look in the detail at this latter case. We could describe $U^{(1)}+G^{(1)}$ having 19 degenerate codons, collectively, by the term $2^{19}$ in Eq.(3) and $A^{(1)}+C^{(1)}$ having 22 degenerate codons, in some detail, by the remaing terms describing the following groups $L^{IV}+P+R^{IV}$ (9) $S^{II}+R^{II}$ (4), T (3); I (2) and H+Q+N+K (1+1+1+1). Conversely, as in case (i) and using the same permutation in (3), we could describe $A^{(1)}+C^{(1)}$ having 22=19+3 degenerate codons, collectively, by the term $2^{19} \times 7^3$, and $U^{(1)}+G^{(1)}$ having 19 degenerate codons, in some detail, by the remaining terms describing the groups V+A+G (9), $S^{IV}+F$ (4), $L^{II}$ (2) and D+E+C+Y (1+1+1+1). We see therefore that the sequence (2) constitutes a robust model which is protected by symmetry. The resulting Gödel Number in Eq.(4) leads, as we have shown in great detail (Négadi, 2007, 2008) to the correct multiplet structure of the standard genetic code by associating the 20 amino acids and the 3 stops to its 23 decimal digits as follows:

5 "quartets": {3, 5, 5, 7, 7}
9 "doublets": {4, 4, 6, 6, 6, 8, 8, 8, 8}
3 "sextets":  {1, 2, 9}                                              (5)

1 "triplet": {2}
2 "singlets": {0, 0}
3 "stops": {0, 0, 0}

Also, it has been shown that the degeneracies of the five amino acids multiplet classes could be easily computed from the decimal digits in each class. By denoting $\nu_j$ and $\sigma_j$ (j=1,2,3,4,6) respectively the number of digits and the sum of their values, without repetition, in each multiplet class j, we have

$\nu_4+\sigma_4 =5+15=20$
$\nu_2+\sigma_2/2 =9+9=18$
$2\nu_6+\sigma_6 =3+(3+12)$ or $(3+3)+12=18$
$\nu_3+\sigma_3 =1+2=3$ (6)
$\nu_1+\sigma_1 =2+0=2$
$\nu_{stops}+\sigma_{stops} =3+0=3$

The special symbol zero was attributed, first, to the two singlets methionine and tryptophane as their degeneracy is zero and, second, to the three stops, as they correspond to the "absence" of amino acids. Note that in Eq.(6) the classes 1, 3 and 4 are perfectly described without any manipulation. As for the doublets and the sextets, only a slight intervention is necessary. For the former, which belong to the only class for which the number of amino acids is strictly equal to the total degeneracy in that class, we halve $\sigma_2$. For the latter (including the three amino acids serine, leucine and arginine) which appear twice, each as a quartet and a doublet, we double $\nu_6$, and put the result in two ways (see Eq.(6). In Summary, we could write either

20+41+3=64 (7)
or
23+38+3=64 (7)'

These last two relations describe nicely the genetic code strucure by emphazising the role of the different numbers in the two "views" mentioned in the introduction and compare very well with the corresponding numbers computed directly from the experimental data in Table 1. In Eq.(7) we have 20 amino acids (aas), 41 degenerate codons, i.e., 5×3+9×1+3×(3+2)+1×2+2×0 and 3 stops. In Eq.(7)' we have 23 amino acids signals (AASs), 38 degenerate codons, i.e., 5×3+9×1+3×(3+1)+1×2+2×0 and also 3 stops.

Degeneracy and chemical composition
Now, we move to the second part of this paper, that is the connection between the genetic code degeneracy and our arithmetical model based on Gödel encoding, on the one hand, and the chemical (atomic) composition of the 20 canonical amino acids, on the other.

These latter are small molecules, the building-blocks of proteins and are all composed of only five types of atoms hydrogen, carbon, nitrogen, oxygen and sulfur. The total number of these atoms in the 20 amino acids are respectively 117, 67, 9, 9 and 2 (see Table 2).

| M | amino acid | hydrogen | carbon | N/O/S | atom | nucleon |
|---|---|---|---|---|---|---|
| 4 | Proline (P) | 5 | 3 | 0 | 8 | 41 |
|   | Alanine (A) | 3 | 1 | 0 | 4 | 15 |
|   | Threonine (T) | 5 | 2 | 0/1/0 | 8 | 45 |
|   | Valine (V) | 7 | 3 | 0 | 10 | 43 |
| 6 | Glycine (G) | 1 | 0 | 0 | 1 | 1 |
|   | Serine (S) | 3 | 1 | 0/1/0 | 5 | 31 |
|   | Leucine (L) | 9 | 4 | 0 | 13 | 57 |
|   | Arginine (R) | 10 | 4 | 3/0/0 | 17 | 100 |
| 2 | Phenylalanine (F) | 7 | 7 | 0 | 14 | 91 |
|   | Tyrosine (Y) | 7 | 7 | 0/1/0 | 15 | 107 |
|   | Cysteine (C) | 3 | 1 | 0/0/1 | 5 | 47 |
|   | Histidine (H) | 5 | 4 | 2/0/0 | 11 | 81 |
|   | Glutamine (Q) | 6 | 3 | 1/1/0 | 11 | 72 |
|   | Asparagine (N) | 4 | 2 | 1/1/0 | 8 | 58 |
|   | Lysine (K) | 10 | 4 | 1/0/0 | 15 | 72 |
|   | Aspartic acid (D) | 3 | 2 | 0/2/0 | 7 | 59 |
|   | Glutamic acid (E) | 5 | 3 | 0/2/0 | 10 | 73 |
| 3 | İsoleucine (I) | 9 | 4 | 0 | 13 | 57 |
| 1 | Methionine (M) | 7 | 3 | 0/0/1 | 11 | 75 |
|   | Tryptophane (W) | 8 | 9 | 1/0/0 | 18 | 130 |
|   | Total | 117 | 67 | 9/9/2=20 | 204 | 1255 |

Table 2. Hydrogen, carbon, N/O/S, atom and nucleon numbers of the 20 amino acids
(side-chains)

Note that here the number of hydrogen atoms, 117, corresponds to the case where in proline, a special amino acid (more precisely an imino acid) with an ill-defined side-chain (see shCherbak, 2003), and we take 41 as the number of nucleons in its side-chain which implies 74 nucleons in its block, see below. In what follows, we shall consider not only 20 amino acids or, as we have seen above 23 amino acids signals (AASs), but also and interestingly 61 "amino acids" corresponding to the 61 meaningful codons. It has been shown by shCherbak (shCherbak, 2003) and Rakočević (Rakočević, 2004) that doing so could lead to the existence of many remarkable mathematical balances involving nucleon and atom numbers. We shall further exploit this path here and derive also remarkable results. Let us first begin by considering hydrogen. In the 61 "amino acids" there are 358 hydrogen atoms (21 in the 5 quartets, 22 in the 3 sextets, 50 in the 9 doublets, 9 in the triplet and 15 in the two singlets, so that 21×4+22×6+50×2+9×3+15×1=358. Taking the sum of the prime factors of this number (358=2×179), and focusing on the prime indices of these latter and their sum, we have derived what we called shCherbak's proline identity (Négadi, 2008):

$$42=41+1$$
or  (8)
$$41=42-1$$

Now, separating the 20 amino acids from the 41 "others", corresponding to the 41 degenerate codons, we have 117 in the former set and 241 (358-117) in the latter. It is remarkable to find that these last two numbers are easily obtained from our Gödel encoding equations (3) and (4). As a matter of fact we have, for the latter, using the functions $a_0$ and $\Omega$ defined in the introduction:

$$a_0(23!)+\Omega(23!)=200+41=241 \tag{9}$$

As for the former, 117, it is simply obtained, as the sum of the number of digits (other than 0), i.e. 18, and the sum of their values, 99, from the decimal representation of 23! (see Eq.(4)). We have therefore 117+241=358, matching exactly the hydrogen pattern, the first from the decimal representation and the second from the prime factorization.

Next, we consider the carbon composition and its relation to degeneracy as derived from our arithmetic equations. As we have seen above (see Eq.(7)' in section 1) the consideration of 23 objects could also lead to interesting results. For example numbering all 23 objects from 1 to 23 and computing the sums for the "singlets", on the one hand, and the "doublets", on the other, we have

$$\sum_{i=1}^{17} i = 153, \quad \sum_{i=18}^{23} i = 123 \tag{10}$$

Application of the arithmetic functions to the above sums gives

$$a_0(153)+a_0(123)=3+3+3+17+41=67 \tag{11}$$

and

$$a_1(153)+a_1(123)=3+3+17+41=20+41+3=64 \tag{12}$$

These last relations are very interesting and have nice interpretations. The first, for example, written as 17+(3+3)+41+3, could be interpreted as follows: 17 amino acids with no first-base degeneracy, 3 sextets in the form ($S^{IV}$, $L^{IV}$, $R^{IV}$)+( $S^{II}$, $L^{II}$, $R^{II}$), 41 degenerate codons and finally 3 stops. Leaving aside the degenerate codons, we have therefore 23 AASs and 3 stops, i.e., 26 objects. This same relation have also another possible interpretation in terms of the total carbon atom number in the 20 amino acids side-chains which is 67. We have shown recently (Négadi, 2007, 2008) how this number appears from our arithmetical model in the form 67=25+42 (there are 25 carbon atoms in classes 1, 3 and 4, and 42 carbon atoms in classes 2 and 6). In the present case, we could write the sum in Eq.(11) as 26+41. To show that these two last patterns do indeed match it is sufficient to make appeal to proline's identity (Eq.(8)) and introduce it in either one of them. Now, the second sum gives the total number of codons 64, partitioned correctly: 20 amino acids (17 and 3 sextets S, L and R), 41 degenerate codons and 3 stops. Still concerning carbon, we consider a fundamental or primitive symmetry residing in the doublets discovered by Findley and his collaborators (Findley, Findley and MacGlynn, 1982). These authors established a one-to-one correspondence from one member of a doubly degenerate codon pair to the other member. Now, we call up the above fundamentalness of the degeneracy of order 2 and consider in our equation (5) the 9 "doublets" and the sum of all nine digits. These, in themselves, lead us to interesting results in connection with carbon patterns. As a matter of fact we have

$$\nu_2+(4+4+6+6+6+8+8+8+8)=9+(29+29)=38+29=67 \qquad (13)$$

First, this sum is very interesting as it gives the exact number of carbon atoms in the 20 amino acids side-chains. Moreover, this number is in full agreement with Petoukhov's classification of the 20 amino acids into the two categories of Complementary (G, P, K, F, A, R, M, Y) and Non-Complementary (V, L, T, S, D, H, I, C, E, Q, N, W) amino acids. In the former there are pairs of amino acids for which the codons that code for the first member of the pair are anti-codons of those codons that code for the second member of the pair (see Petoukhov, 2001). In the latter such correspondence does not exist. It is seen that there are 29 carbon atoms in the eight Complementary amino acids and 38 (=9+29) in the twelve Non-Complementary amino acids (see Tables 1 and 3).

| C aas | | NC aas | | |
|---|---|---|---|---|
| Glycine | Alanine | Valine | Aspartic acid | Glutamic acid |
| Proline | arginine | Leucine | Histidine | Glutamine |
| Lysine | Methionine | Threonine | İsoleucine | Asparagine |
| phenylalanine | tyrosine | serine | cysteine | tryptophane |
| 29 carbon atoms | | 38 carbon atoms | | |

Table 3. Petoukhov's classification into Complementary and Non-Complementary amino acids

Note interestingly that this is also in good agreement with the recent classification of the 20 amino acids by Štambuk and his collaborators (Štambuk, Konjedova and Gotovac, 2004) into polar and nonpolar amino acids using the "scale based agglomerative clustering method".

From the above reference the nonpolar group includes the amino acids Y, M, V, L, F, I, W and C, and the polar group includes the following ones H, R, Q, K, N, E, D, P, A, T, S and G. We find that the first group has 38 carbon atoms and the second 29. Also, the above sum in the form 9+58 corresponds to another carbon atom pattern found by us in connection with the nucleon composition patterns (see Négadi 2006), and the 28-gon classification by Yang (Yang, 2004) where his pyramid-shaped "primordial codon core" composed of the five amino acids glycine, proline, alanine, valine and threonine has 9 carbon atoms, leaving 58 for the remaining amino acids. Moreover, taking $2 \times v_2$ in place of $v_2$, by virtue of the one-to-one correspondence, reveals, first, the number of carbon atoms in the 23 AASs 9+67=76 and, second, other interesting carbon atom patterns in the genetic code table. We could write either

$$(9+9+29)+29=47+29 \tag{14}$$
or
$$(9+29)+(9+29)=38+38 \tag{14'}$$
or
$$(9+9)+(29+29)=18+58 \tag{14''}$$

First, from (14), we have that there are 47 carbon atoms in $U^{(1)}+C^{(1)}$ and 29 carbon atoms in $A^{(1)}+G^{(1)}$. Second, there are 38 carbon atoms in $U^{(1)}+G^{(1)}$ and 38 carbon atoms in $A^{(1)}+C^{(1)}$, a nice balance, in Eq.(14)'. Note that these last two cases are in full agreement with the two partitions, into two moieties, and leading to the starting sequence in Eq.(2) in our arithmetic model based on Gödel encoding (see section 1). Third, Eq.(14)'' gives the number of carbon atoms in the well known Rumer sets $M_1$ and $M_2$, called also four-codons and non-four-codons, respectively (see for example shCherbak, 2003 and Rakočević, 2008). Recall that in $M_1$ (in grey in Table 1) with 32 codons there are eight quartets of codons each of which codes for a single amino acid, that is the 5 true quartets P, A, G, T, V and the 3 quartet-part of the 3 sextets $S^{IV}$, $L^{IV}$, $R^{IV}$) while in $M_2$, also with 32 codons and eight quartets, the codons of which, now, code for more than one amino acid, i.e., the 9 doublets F, Y, C, H, Q, N, K, D, E, the 3 doublet-part of the 3 sextets $S^{II}$, $L^{II}$, $R^{II}$, the triplet I and finally the 2 singlets M and W. These two sets are exchanged by the Rumer transformation (see above). As a result, we have that there are 18 carbon atoms in $M_1$ and 58 in $M_2$, as in Eq.(14)''.

Now, we consider atom numbers. From Table 2, we have 204 atoms in the 20 amino acids side-chains. Evaluating the number of atoms in the 61 "amino acids" (side-chains), as above for hydrogen, we have 31 in the quartets, 35 in the sextets, 96 in the doublets, 13 in the triplet and finally 29 in the singlets so that their total number is 594. The number of atoms corresponding to the 41 degenerate codons is 390 (=594-204). In view of the numeric importance of hydrogen and carbon with respect to nitrogen, oxygen and sulfur (the former two atoms compose slightly more than 90% of the total), we make the following partition 204=184+20

(CH/NOS), use the functions $a_0(390)=23$, $a_0(184)=29$, $a_0(20)=9$, and write either

$$a_0(390)+[a_0(184)+a_0(20)]=23+(29+9)=23+38=61 \qquad (15)$$
or
$$[a_0(390)+ a_0(20)]+ a_0(184)=(23+9)+(29)=32+29=61 \qquad (15)'$$

Here we have that Eq.(15) conforms to our Eq.(7)': 23 AASs and 38 degenerate codons according to the second "view" (see the introduction). As for Eq.(15)' it gives exactly the number of codons in $U^{(1)}+C^{(1)}$ (29) and in $A^{(1)}+G^{(1)}$ (32), on the one hand, and in $U^{(1)}+G^{(1)}$ (29) and in $A^{(1)}+C^{(1)}$ (32), on the other. This is also in nice agreement with the two chosen partitions in our arithmetic model based on Gödel encoding (see section 1).

We consider finally the nucleon numbers. In 61 "amino acids" side-chains, there are 3404 nucleons ($145\times4+660\times2+188\times6+57\times3+75+130$), see Table 1). We have immediately $a_0(3404)=23+(2+2+37)=23+41=64$. This is nothing but what our Eq.(7) says: 20 amino acids, 3 stops and 41 degenerate codons.

Conclusion

In this short paper, we have demonstrated a clear connection between the (codon) degeneracy structure of the standard genetic code, as described by a new more flexible symmetry-based version of our arithmentical model using the mathematical technique of Gödel encoding, and the chemical composition of the twenty canonical amino acids. This latter include the hydrogen, carbon, atom and nucleon contents. As new results, in section 2, we have derived several remarkable patterns in full agreement with several recent published works on the genetic code. Although this study shows no obvious practical application(s) we nevertheless hope that it could add, together with the results obtained earlier (Négadi, 2007, 2008), some new knowledge in the field and help further understanding of the genetic code physico-mathematical structure.